\documentclass[prd, 10pt,aps,nofootinbib,floatfix,notitlepage,twocolumn,superscriptaddress]{revtex4-2}
\usepackage[margin=1in]{geometry}
\usepackage{amsmath,amssymb}
\usepackage{mathtools}
\usepackage{graphicx}
\usepackage{booktabs}
\usepackage{multirow}
\usepackage[colorlinks=true,linkcolor=blue,citecolor=blue,urlcolor=blue]{hyperref}

\newcommand{\AP}{Alcock--Paczy\'nski\ }

\newcommand{\rd}{r_\mathrm{d}}
\newcommand{\Fap}{F_{\rm AP}}
\renewcommand{\max}{\mathrm{max}}

\newcommand{\Cov}{\mathrm{Cov}}
\renewcommand{\d}{\mathrm{d}}
\newcommand{\e}{\mathrm{e}}
\newcommand{\avg}{\mathrm{avg}}
\newcommand{\zref}{z_\mathrm{ref}}
\newcommand{\true}{\mathrm{true}}
\newcommand{\std}{\mathrm{std}}
\newcommand{\corr}{\mathrm{corr}}
\newcommand{\obs}{\mathrm{obs}}
\newcommand{\coast}{\mathrm{coast}}

\begin{document}
\title{Model-independent and reconstruction-free evidence for cosmic acceleration from DESI baryon acoustic oscillations}

\author{Laura Herold}
\email{lherold@cern.ch}
\affiliation{William H. Miller III Department of Physics and Astronomy, Johns Hopkins University, 3400 North Charles Street, Baltimore, MD 21218, USA}
\author{Graeme E.\ Addison}
\email{gaddison@jhu.edu}
\affiliation{William H. Miller III Department of Physics and Astronomy, Johns Hopkins University, 3400 North Charles Street, Baltimore, MD 21218, USA}
\author{Adam G.\ Riess}
\email{ariess@stsci.edu}
\affiliation{William H. Miller III Department of Physics and Astronomy, Johns Hopkins University, 3400 North Charles Street, Baltimore, MD 21218, USA}
\affiliation{Space Telescope Science Institute, 3700 San Martin Drive, Baltimore, MD 21218, USA}

\begin{abstract}
    While there is ample evidence that the universe's expansion is accelerating, most of this evidence relies on assuming a shape of the relative expansion rate, $H(z)/H_0$, informed by a cosmological model or a kinematical approach. Model-independent analyses are common for supernova datasets but rarely applied to other data. In this work, we use uncalibrated baryon acoustic oscillations (BAO) from DESI DR2 for a model-independent test of acceleration, which makes no assumptions about the theory of gravity or the energy content of the universe. We test for cosmic acceleration with three model-independent statistics that rely purely on the measured Alcock--Paczy\'nski parameter, the isotropic BAO, and the line-of-sight BAO, and do not require any reconstruction between redshift bins unlike previous analyses. Accounting for the look-elsewhere effect, among all statistics we find $2.4\sigma$ -- $4.1\sigma$ evidence for cosmic acceleration within $0 \leq z \leq 0.934$ when assuming flatness; allowing for spatial curvature, $|\Omega_k| \leq 1$, at least two of the three statistics retain $2.4\sigma$ -- $3.5\sigma$.
\end{abstract}

\maketitle

\section{Introduction}
\label{sec:intro}

\begin{figure}[t]
    \centering
    \includegraphics[width=\columnwidth]{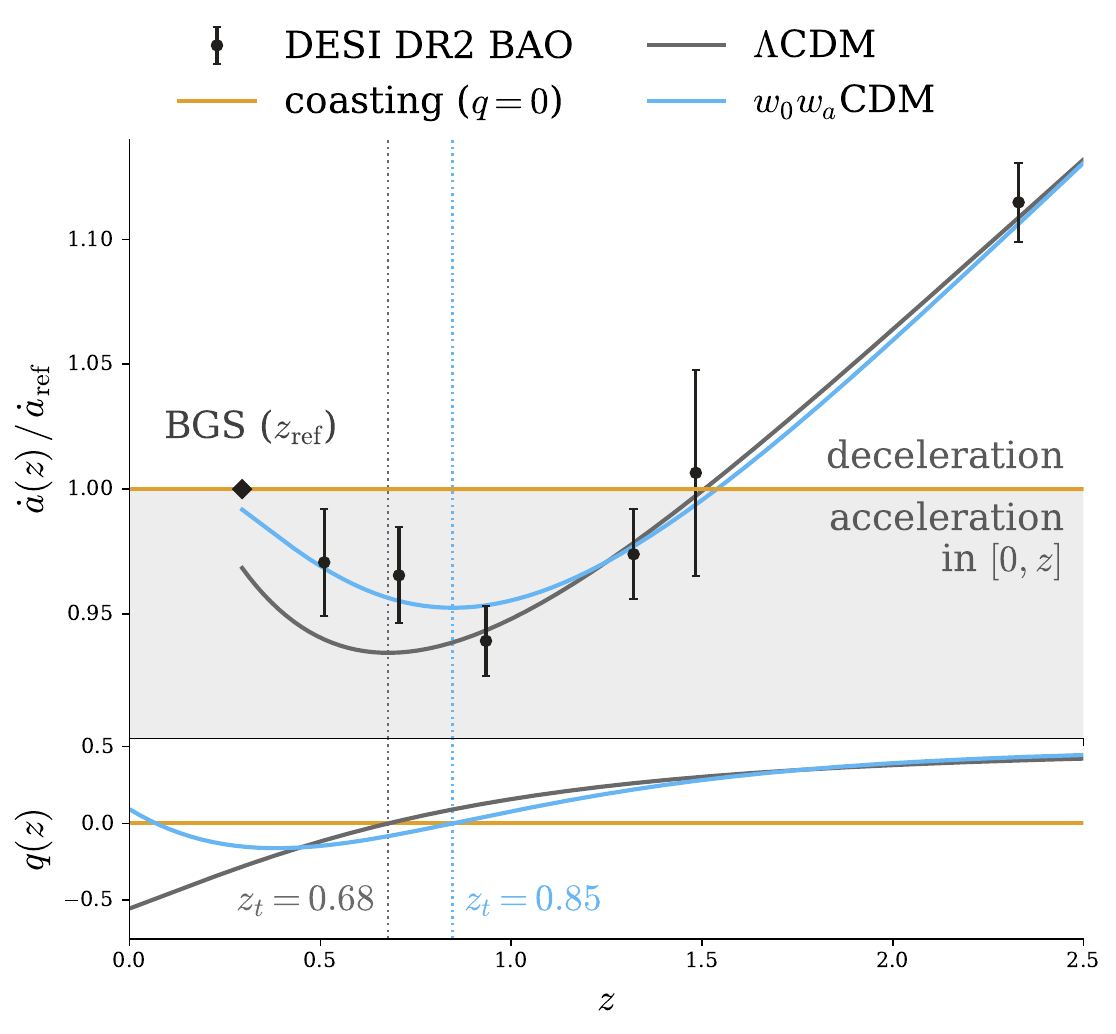}
    \caption{\textit{Top:} BAO (markers) constrain the relative expansion rate $E(z)$ and thus the expansion speed, $\dot a$, normalized to the effective coasting speed, $\dot{a}_\mathrm{ref}$, inferred from the bright galaxy sample (BGS) at $\zref = 0.295$ (Sec.~\ref{sec:methods} for details). DESI DR2 data lie on top of accelerating models ($\Lambda$CDM, $w_0w_a$CDM; Sec.~\ref{sec:data}). Data with $\dot a (z) < \dot{a}_\mathrm{ref}$ (gray shaded area) indicate acceleration within [0, $z$].
    \textit{Bottom:} Deceleration parameter $q(z)$ of the three models; vertical dotted lines mark the transition, $z_t$, from deceleration to acceleration.}
    \label{fig:dot_a}
\end{figure}

In 1998, observations of type Ia supernovae (SNe) obtained evidence for an accelerated expansion of the universe  \citep{SupernovaSearchTeam:1998fmf,SupernovaCosmologyProject:1998vns}, i.e.\ for a negative deceleration parameter
\begin{equation}
    \label{eq:q(t)}
	q(t) = -\frac{\ddot{a}}{aH^2} <0,
\end{equation}
where $a$ is the scale factor, $H$ the Hubble parameter, and `dot' refers to derivatives with respect to cosmological time $t$.  Within general relativity, an accelerating universe requires a component with an equation of state $w<-1/3$, where the simplest such component is a cosmological constant $\Lambda$. While there is significant model-dependent evidence for the existence of such a component from the cosmic microwave background (CMB) and baryon acoustic oscillations (BAO), the large majority of model-independent measurements of $q(t)$ rely on SNe. 

Cosmological models favored by current CMB and BAO data predict a transition from deceleration to acceleration around a redshift, $z \approx 0.6$ -- $0.9$, illustrated in Fig.~\ref{fig:dot_a}. Thus SN data, which typically cover $0 < z < 1$, are a sensitive probe of acceleration. Most tests for acceleration based on SNe assume a shape of the relative expansion rate 
\begin{equation}
    E(z) = \frac{H(z)}{H_0},
\end{equation}
often based on a cosmological model, for example $\Lambda$ cold dark matter ($\Lambda$CDM). Within flat $\Lambda$CDM, $q_0 = -1 + 3\Omega_\mathrm{m}/2$ and the densities of matter and $\Lambda$ add up to unity, $\Omega_\mathrm{m} + \Omega_\Lambda = 1$; thus $\Omega_\mathrm{m} < 2/3$ automatically implies acceleration. An alternative is to expand  $E(z)$ as a kinematic (Taylor) series, written in terms of $a$ and its time derivatives, including $q$ \cite{Visser:2003vq, Turner:2001mx, SupernovaSearchTeam:2004lze, Shapiro:2005nz}. The latter approach avoids a dependence on the cosmological model, however, being an expansion around $z=0$, it is known that the series does not converge for $z>1$ \cite{Cattoen:2007sk}. To bypass any assumptions about the shape of $E(z)$, Refs.~\cite{Turner:2001mx, Visser:1997tq, Gong:2007zf, Seikel:2007pk, Seikel:2008ms, Yang:2019fjt} suggest a purely geometric, model-independent test for acceleration based only on the luminosity distance, $D_L$, of individual SNe. More recent works~\cite{Zunckel:2008ti, Holsclaw:2010sk, Seikel:2012uu, Tutusaus:2018ulu} extend this idea by reconstructing\footnote{In this work, ``reconstruction'' refers to reconstructing the functional shape and its uncertainty in redshift space, not density-field reconstruction applied to sharpen the BAO measurements.} the full functional shape of quantities, e.g.\ $D_L(z)$, and their derivatives using reconstruction algorithms like Gaussian process reconstruction (GPR) in order to increase the sensitivity to acceleration. 

While SN measurements are a natural choice to constrain acceleration, BAO are a fundamentally different probe of the universe's expansion with different systematic uncertainties. In this work, we use recent BAO data from DESI DR2 \cite{DESI:2025zgx} to estimate the evidence for cosmic acceleration without relying on a cosmological model. We assume flatness in the main text and show in App.~\ref{app:curvature} that allowing for curvature does not change the main conclusions.

Using BAO to constrain acceleration was already suggested by Ref.~\cite{Sutherland:2011uy}, who proposed a model-independent test based on isotropic and line-of-sight BAO measurements similar to the model-independent tests based on SNs' $D_L$. More recent works reconstruct $q(z)$ from BAO measurements using GPR and other reconstruction methods, both pre-DESI \cite{Gomez-Valent:2018gvm, Haridasu:2018gqm, Yu:2017iju} and post-DESI \citep{Mukherjee:2024ryz, Gonzalez-Fuentes:2025lei, DESI:2025fii, Hu:2025fsz, Lu:2024hvv, Yang:2025cgx}. However, Refs.\ \cite{Hwang:2022hla, Johnson:2025blf, Ruchika:2025mkx} pointed out that the results depend sensitively on the reconstruction method used. 

In this work, we use three model-independent reconstruction-free statistics to probe cosmic acceleration from BAO. One of these statistics was already proposed by Sutherland in 2011 \cite{Sutherland:2011uy} but -- to our knowledge -- never applied to real data. After describing the data and simulations in Sec.~\ref{sec:data}, we define these three statistics in Sec.~\ref{sec:methods}, we present our results in Sec.~\ref{sec:results} and conclude in Sec.~\ref{sec:conclusions}.


\section{Data and simulations}
\label{sec:data}

\begin{figure*}[t]
    \centering
    \includegraphics[width=1\textwidth]{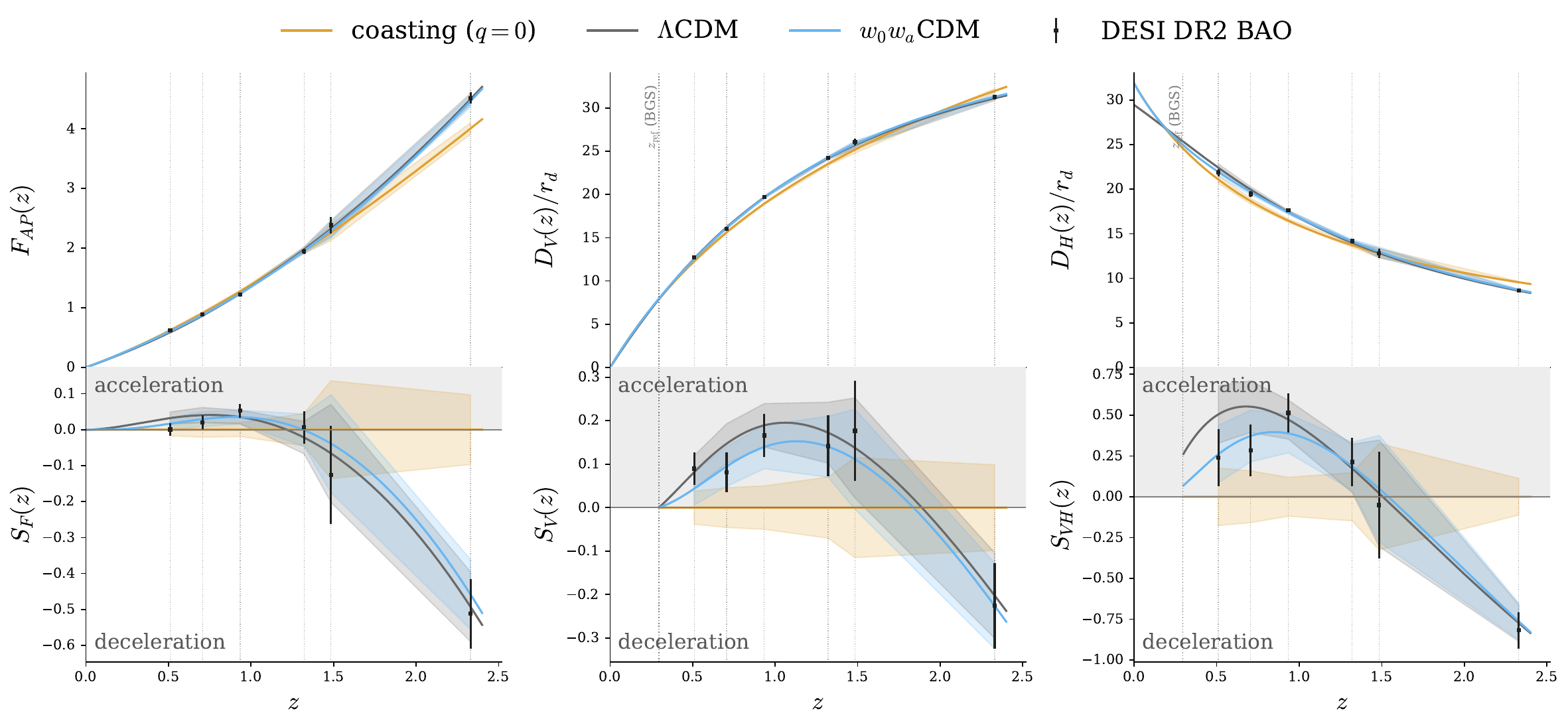}
    \caption{\textit{Top:} DESI DR2 measurements (markers) of the \AP  parameter, $\Fap(z)$ (left), isotropic BAO, $D_V/\rd$ (center), and line-of-sight BAO, $D_H/\rd$ (right), along with three cosmological models: a coasting cosmology ($q(z)=0$), $w_0w_a$CDM (posterior mean of DESI+CMB), and $\Lambda$CDM (posterior mean of DESI) with 68\%-C.L.\ bands from MC simulations.
    \textit{Bottom:} Model-independent and reconstruction-free statistical tests to quantify acceleration: $S_F(z)> 0$ (left), $S_V(z) > 0$ (center), or $S_{VH}(z) > 0$ (right) for any $z$ imply acceleration somewhere in $[0,\, z]$.}
    \label{fig:SF_SV}
\end{figure*}

Line-of-sight BAO are sensitive to the ratio of the comoving Hubble distance $D_H$ and the comoving size of the sound horizon $\rd$, while transverse BAO are sensitive to the ratio of the comoving angular diameter distance $D_M$ and $\rd$, i.e.\ for a flat universe:
\begin{equation}
    \label{eq:DH_DM}
    \frac{D_H(z)}{\rd} = \frac{c}{\rd\,H(z)}, \quad
    \frac{D_M(z)}{\rd} =\frac{c}{\rd}\int_0^z \frac{\d z'}{H(z')},
\end{equation}
where $c$ is the speed of light. The isotropic BAO scale is given by the angle-averaged quantity \cite{SDSS:2005xqv}
\begin{equation}
    \label{eq:DV}
    D_V(z)=\big[z\,D_H(z)\,D_M(z)^2\big]^{1/3}.
\end{equation}
It is common to define the \AP parameter as the ratio of the two BAO distances,
\begin{equation}
    \label{eq:FAP}
    \Fap(z) = \frac{D_M(z)}{D_H(z)},
\end{equation}
such that the dependence on $r_\d$ and $H_0$ drops out. For completeness, we quote the DESI DR2 measurements of $\Fap$, $D_V/\rd$, and $D_H/\rd$ used in this work in Tab.~\ref{tab:DESI_data} (taken from \cite{DESI:2025zgx})\footnote{The DESI collaboration samples the BAO posterior in the isotropic and anisotropic scaling parameters, $\alpha_{\rm iso}\propto D_V/\rd$ and $\alpha_{\rm AP}\propto\Fap^{-1}$, and quotes the marginalized mean and standard deviation of each distance. We treat these marginals as Gaussian and different redshift bins as uncorrelated following the DESI convention.} and show them as black markers in the top panels of Fig.~\ref{fig:SF_SV}. In the following we will refer to the DESI DR2 BAO measurements as ``DESI''.
\begin{table}[h]
  \centering
  \resizebox{\columnwidth}{!}{%
  \begin{tabular}{lcccc}
  \toprule
Tracer & $z_{\mathrm{eff}}$ & $\Fap$ & $D_V/\rd$ & $D_H/\rd$ \\
\midrule
BGS       & 0.295 & ---              & $7.942\pm0.075$  & --- \\
LRG1      & 0.510 & $0.622\pm0.017$  & $12.720\pm0.099$ & $21.863\pm0.425$ \\
LRG2      & 0.706 & $0.892\pm0.021$  & $16.050\pm0.110$ & $19.455\pm0.330$ \\
LRG3\\
\ +ELG1 & 0.934 & $1.223\pm0.019$  & $19.721\pm0.091$ & $17.641\pm0.193$ \\
ELG2      & 1.321 & $1.948\pm0.045$  & $24.252\pm0.174$ & $14.176\pm0.221$ \\
QSO       & 1.484 & $2.386\pm0.136$  & $26.055\pm0.398$ & $12.817\pm0.516$ \\
Lyman-$\alpha$    & 2.330 & $4.518\pm0.097$  & $31.267\pm0.256$ & $8.632\pm0.101$ \\
\bottomrule
\end{tabular}}
  \caption{DESI DR2 \AP parameter, $\Fap$, isotropic BAO, $D_V/\rd$, and line-of-sight BAO, $D_H/\rd$, used in this work (taken from Table IV in \cite{DESI:2025zgx}).}
  \label{tab:DESI_data}
\end{table}

We generate Monte-Carlo (MC) simulations to test our pipeline and inform choices within the definition of the statistical tests (Sec.~\ref{sec:results}). We use three cosmological models for the simulations, but emphasize that the choice of these particular models does not impact the quoted significances from the real data. The models are:

\begin{itemize}
    \item[-] ``Coasting'': a universe with $q(z) = 0$ for all $z$;
    \item[-] ``$\Lambda$CDM'': the DESI posterior mean under the $\Lambda$CDM model of the matter density, $\bar\Omega_\mathrm{m} = 0.2975$ (Tab.~5 in \citep{DESI:2025zgx});
    \item[-] ``$w_0w_a$CDM'': the posterior means under DESI, Planck \cite{Planck:2018vyg}, and Atacama Cosmology Telescope lensing \cite{ACT:2023dou} within the Chevallier-Polarski-Linder parameterization \cite{Chevallier:2000qy, Linder:2002et}: $\bar\Omega_\mathrm{m} = 0.353$, $\bar{w_0} = -0.42$, $\bar{w_a} = -1.75$ (Tab.~5 in \citep{DESI:2025zgx}).
\end{itemize}

These parameters uniquely determine the relative expansion rate via $E(z) = 1+z$ (coasting) or via $E(z) = \sqrt{(1+z)^3\,\Omega_\mathrm{m} + (1+z)^{\nu}\,\mathrm{e}^{-3 w_a z/(1+z)}(1-\Omega_\mathrm{m})}$ 
with $\nu \equiv 3(1+w_0 + w_a)$ ($\Lambda$CDM, $w_0w_a$CDM). We analytically fit the overall amplitude $H_0\, r_\d$ of the ground truth of each theory model to DESI $D_H/r_\d$ data. This sets the theory values of the statistics defined in Sec.~\ref{sec:methods} for the simulations but does not affect any statistic or significance inferred from real DESI data. 

Using the respective ground truth of each theory model as mean and the DESI-quoted errors as standard deviation (Tab.~\ref{tab:DESI_data}), we draw $10^7$ Gaussian realizations of $\Fap$ and $D_H/\rd$ at six DESI redshifts and of $D_V/\rd$ at seven DESI redshifts.


\section{Reconstruction-free tests of cosmic acceleration}
\label{sec:methods}

Our goal is to quantify evidence for a negative deceleration parameter, $q(z) < 0$ (Eq.~\ref{eq:q(t)}), from BAO alone. One can express $q(z)$ in terms of the relative expansion rate:
\begin{equation}
    q(z) = \frac{E'(z)}{E(z)}(1+z) - 1. 
\end{equation}
This is a differential equation for $E(z)$, i.e.\ $\d \ln E(z) / \d z = (q(z) +1) / (1+z)$, with the solution 
\begin{equation}
    \label{eq:E(z)_solution}
    E(z) = (1+z)\, \e^{Q(z)};\quad Q(z) \equiv \int_0^z \frac{q(z')}{1+z'}\d z'.
\end{equation}
Thus if the universe never experienced any acceleration, $q(z) \ge 0\ \forall z$, this implies (a) $Q(z) \geq 0$ for all $z$ and (b) $Q(z)$ never-decreasing with $z$. In the following, we assume a never-accelerating universe as the null hypothesis and show that this hypothesis is violated.
We define model-independent and reconstruction-free statistics based on $\Fap(z)$, on $D_V(z)/D_V(\zref)$ and on $D_H(z)/D_V(\zref)$.\footnote{We discuss in Sec.~\ref{sec:discussion} why these statistics are more sensitive than others.}

The statistics are based solely on the Friedmann-Lema\^{\i}tre-Robertson-Walker metric, i.e.\ they assume statistical homogeneity and isotropy; they do not depend on the theory of gravity (e.g.\ general relativity) or on the energy content of the universe, i.e.\ they are independent of the Friedmann equations. We assume spatial flatness in the main text but show that our main conclusions hold for curved cosmologies in App.~\ref{app:curvature}. $S_V$ and $S_{VH}$ further rely on BAO being a standard ruler, i.e.\ that the BAO feature at $\zref$ and at $z$ corresponds to the same comoving scale, $\rd$, while $S_F$ is independent of this assumption.


\subsection{$\Fap$-based statistic}
\label{sec:S_F}

The \AP parameter, $\Fap$, can be written in terms of $Q(z)$ by inserting Eq.~\ref{eq:DH_DM} and \ref{eq:E(z)_solution} into Eq.~\ref{eq:FAP}:
\begin{equation}
    \label{eq:Fap_Q(z)}
    \Fap(z) = (1+z)\ \e^{Q(z)} \int_0^z \frac{\e^{-Q(z')}}{1+z'} \d z'. 
\end{equation}
Thus for a coasting cosmology with $q(z) = 0$,  $Q(z) = 0$ and the integral can be solved analytically such that $\Fap^\coast = (1+z) \ln(1+z)$. If the universe never underwent any phase of acceleration, i.e.\ $q(z) \ge 0\ \forall z$, then since $Q(z)$ is never-decreasing, $\e^{-Q(z)}$ is never increasing and we can replace the argument of $Q(z)$ with its upper limit and solve the integral (\hypertarget{star}{$\star$}): 
$\int_0^z \e^{-Q(z')}/({1+z'}) \d z' \geq \e^{-Q(z)}\ln(1+z)$. Inserting this in Eq.~\ref{eq:Fap_Q(z)}, a never-accelerating universe satisfies $\Fap \ge (1+z) \ln(1+z)$. We define the statistic:
\begin{equation}
    \label{eq:S_F}
    S_F(z) \equiv (1+z)\ln(1+z)- \Fap(z).
\end{equation}
For a never-accelerating universe $S_F(z) \leq 0$. Thus if $S_F(z)>0$ for any $z$, this is evidence for acceleration somewhere in the redshift range $[0,z]$. Since $S_F$ is linear in $\Fap$, the uncertainty, $\sigma_F$, of $\Fap$ quoted by the DESI collaboration (Tab.~\ref{tab:DESI_data}) can simply be propagated to $S_F$, i.e.\ $\sigma_{S_F} = \sigma_F$. We show $S_F(z)$ for DESI data in the bottom left panel of Fig.~\ref{fig:SF_SV} along with three theory models.


\subsection{$D_V$-based statistic}
\label{sec:S_V}

Similarly, $D_V$ can be expressed in terms of $Q(z)$ by inserting Eqs.~\ref{eq:DH_DM},\,\ref{eq:E(z)_solution} into Eq.~\ref{eq:DV}:
\begin{equation}
    \label{eq:DV_solution}
    D_V(z) = \frac{c}{H_0} \left[z\frac{\e^{-Q(z)}}{1+z} \Big(\int_0^z\frac{\e^{-Q(z')}}{1+z'}\d z' \Big)^2\right]^{1/3}.
\end{equation}
For $q(z) = 0$, $Q(z) = 0$, this simplifies to $D_V(z) = (c/H_0)\, g(z)$ with
\begin{equation}
    g(z) \equiv \left[\frac{z\, \ln^2(1+z)}{1+z}\right]^{1/3}.
\end{equation}
Note that DESI BAO does not measure $D_V$ directly but $D_V/r_\d$. Moreover, $D_V(z)$ still depends on $H_0$. To avoid a dependence on $H_0\, r_\d$, we compare to a reference redshift, $\zref = 0.295$ (BGS). A never-accelerating universe's $Q(z)$ is never decreasing and thus for $\zref \leq z$: $D_H(z) / D_H(\zref) \leq (1+\zref)/(1+z)$.  
Using (\hyperlink{star}{$\star$}) with $\zref$ as upper limit and 
$\int_{\zref}^{z}e^{-Q}/(1+z')\d z'\ \le\ e^{-Q(\zref)} [\ln(1+z)-\ln(1+\zref)]$, we find $D_M(z)/D_M(\zref) \leq \ln(1+z)/ \ln(1+\zref)$. Therefore, a never-accelerating universe implies $D_V(z)/D_V(\zref) \leq g(z)/g(\zref)$. We use this to define the statistic:
\begin{equation}
    \label{eq:S_V}
    S_V(z) \equiv g(\zref)\,\frac{D_V(z)}{\rd} - g(z)\,\frac{D_V(\zref)}{\rd}.
\end{equation}
Thus $S_V(z) > 0$ for any $z$ implies evidence for acceleration in $[0,\, z]$. We defined $S_V$ such that it is linear in $D_V/\rd$, so the covariance is given by:
\begin{equation}
    \begin{split}
        &\Cov[S_V(z_i), S_V(z_j)] = \\
        &\quad \delta_{ij}\, g^2(\zref)\, \sigma^2_{D_V}(z_i) + g(z_i)\, g(z_j)\, \sigma^2_{D_V}(\zref).
    \end{split}
\end{equation}
We show $S_V(z)$ in the bottom center panel of Fig.~\ref{fig:SF_SV}. Note that while $S_V(z)$ inferred from the DESI $D_V$ measurements (black markers), appear further from zero than $S_F(z)$ inferred from DESI, the $S_V(z)$ measurements at different redshifts are correlated due to the common reference distance $D_V(\zref)$, while the $S_F(z)$ measurements are uncorrelated. We take this effect into account in Sec.~\ref{sec:results}.


\subsection{$D_V/D_H$-based statistic}
\label{sec:S_VH}

Ref.~\cite{Sutherland:2011uy} proposed a test that combines isotropic and line-of-sight BAO at two different redshifts. Inserting Eq.~\ref{eq:E(z)_solution} in Eqs.~\ref{eq:DV}, \ref{eq:DH_DM} yields
\begin{equation}
    \begin{split}
    &\frac{D_V(\zref)}{D_H(z)} = \\
    &(1+z)\left[\frac{\zref}{1+\zref}\right]^{1/3} \!\!\frac{\e^{Q(z)}}{\e^{\frac{Q(\zref)}{3}}}\left(\int_0^{\zref} \!\!\frac{e^{-Q(z')}}{1+z'}\d z'\right)^{2/3}.
    \end{split}
\end{equation}
For a never-accelerating universe and $z>\zref$ we use (\hypertarget{star}{$\star$}) with $\zref$ as upper limit and $e^{Q(z)-Q(\zref)}\ge 1$, which implies $D_V(\zref)/D_H(z) \geq (1+z) g(\zref)$. We can thus define the statistic
\begin{equation}
    \label{eq:S_VH}
    S_{VH}(z) \equiv (1+z)\, g(\zref)\, \frac{D_H(z)}{\rd} - \frac{D_V(\zref)}{\rd},
\end{equation}
where $S_{VH}(z) > 0$ implies acceleration in $[0,\, z]$. We defined $S_{VH}$ such that it is linear in $D_H/\rd$ and $D_V/\rd$, so the covariance is given by:
\begin{equation}
    \begin{split}
        &\Cov[S_{VH}(z_i), S_{VH}(z_j)] = \\
        &\quad \delta_{ij}\, (1+z_i)^2 g^2(\zref)\, \sigma^2_{D_H}(z_i) + \sigma^2_{D_V}(\zref).
    \end{split}
\end{equation}
We show $S_{VH}(z)$ in the bottom right panel of Fig.~\ref{fig:SF_SV}. Similarly to $S_V(z)$, the $S_{VH}(z)$ at different $z$ are correlated due to the shared $D_V(\zref)$.

$D_V(\zref)/D_H(z)$ is directly related to $\dot{a}/\dot{a}_\mathrm{ref}$ shown in Fig.~\ref{fig:dot_a} via
\begin{equation}
    \frac{\dot a}{\dot a_\mathrm{ref}} =\frac{1}{(1+z)g(\zref)}\, \frac{D_V(\zref)}{D_H(z)},
\end{equation}
where $\dot{a} = H/(1+z) =  c /((1+z)D_H)$. To cancel the dependence on $r_\d$, we normalize by the expansion speed of a coasting cosmology at $\zref$ (BGS), which has $\dot a_\mathrm{ref} = H_0 = c\,g(\zref) / D_V(\zref)$. ${\dot a}/{\dot a_\mathrm{ref}} < 1$ in Fig.~\ref{fig:dot_a} thus corresponds to $S_{VH} > 0$.


\begin{figure*}[t]
    \centering
    \includegraphics[width=1\textwidth]{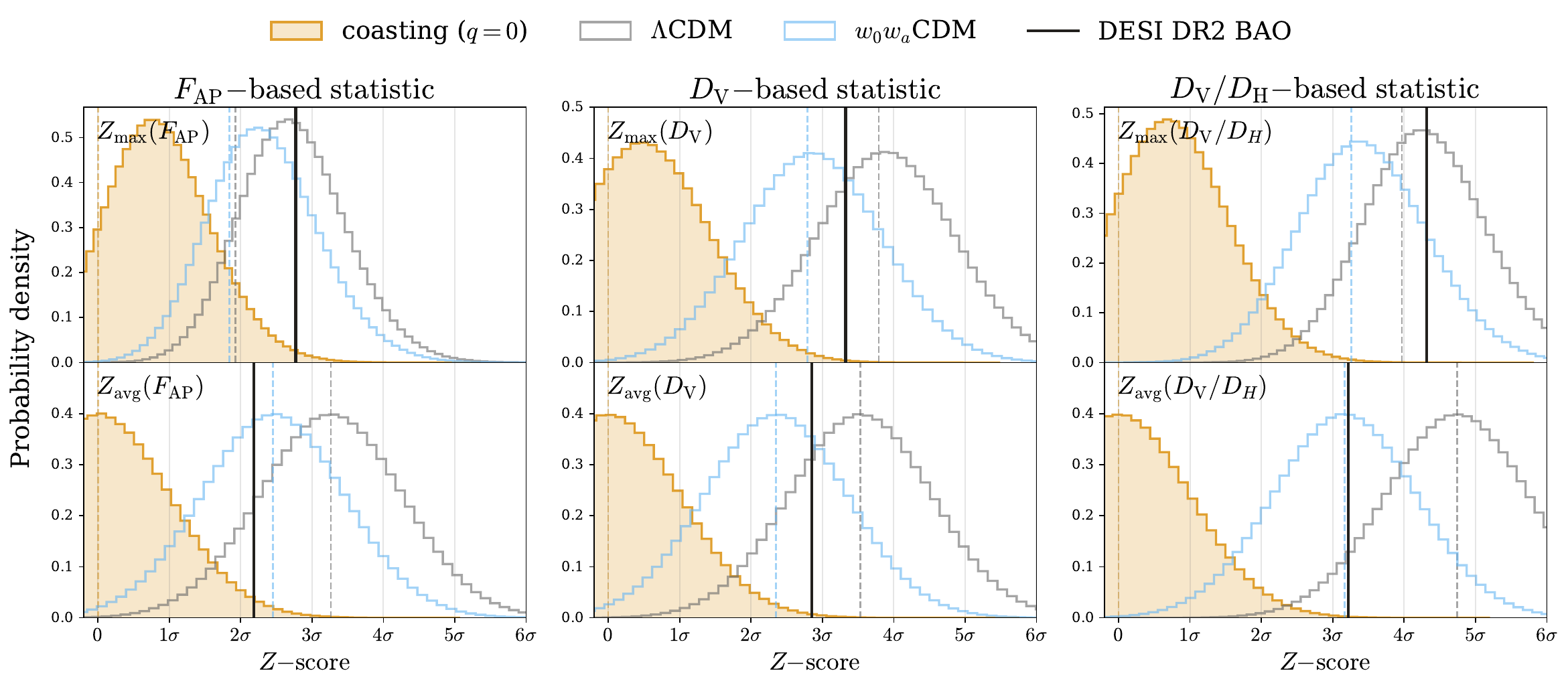}
    \caption{Significance of acceleration as measured by the $Z$-score in units of $\sigma$ for the maximum deviation, $Z_\max$ (top, Eq.~\ref{eq:Z_max}), and the average deviation, $Z_\avg$ (bottom, Eq.~\ref{eq:Z_avg}), for the three statistics $S_F$ (left), $S_V$ (center) and $S_{VH}$ (right). Real DESI data (black solid vertical lines) favor acceleration at $2.2\sigma$ -- $4.3 \sigma$ before the look-elsewhere correction. The distribution of the $10^7$ simulations per theory model (histograms) with the noiseless ground-truth model (dashed vertical lines) are shown for comparison and do not enter $Z$-values of the data.}
    \label{fig:results}
\end{figure*}
\section{Results}
\label{sec:results}

We report evidence for acceleration in two ways: as point-wise maximum significance (Sec.~\ref{sec:Z_max}) and as average significance over multiple redshift bins (Sec.~\ref{sec:Z_avg}) for $S_F$, $S_V$, $S_{VH}$. Note that the different statistical tests are not independent as they share data; we consider all three statistics as a robustness check.
We describe our tests on MC simulations before applying them to DESI data. Notebooks to reproduce the results are publicly available.\footnote{ \url{https://github.com/LauraHerold/BAO-accel}}


\subsection{Point-wise maximum significance, $Z_\max$}
\label{sec:Z_max}

We define the point-wise significance at the redshift that maximizes $S(z)/\sigma_S(z)$ as
\begin{equation}
    \label{eq:Z_max}
    Z_{\rm max} \equiv \max_{i \le m} Z_i\quad \mathrm{with}\quad Z_i = \frac{S(z_i)}{\sigma_S(z_i)}.
\end{equation}
$Z_\max$ is the ``$Z$-score'' that measures the maximum deviation from the non-accelerating universe in units of $\sigma$, over the first $m$ DESI redshift bins (excluding the BGS $D_V/r_\d$ measurement that is used as an anchor for $S_V$ and $S_{VH}$). We pick $m=3$; i.e. $0.510 \leq z \leq 0.934$ as this is the redshift range where acceleration is expected. In App.~\ref{app:redshift}, we illustrate on simulations that this balances higher significance and lower look-elsewhere effect. 

For the $10^7$ MC simulations drawn from the three cosmological models, we show the distribution of $Z_\max$ in the top panels of Fig.~\ref{fig:results}; we quote the mean, $\left< Z_\max \right>$, the ground-truth values, $Z_\true$, and the standard deviation, $\std(Z)$, of the simulations in Tab.~\ref{tab:mock}. The histograms of the MC simulations of the coasting cosmology have a lower mean $Z$-score, $\left<Z_\max\right> \approx 0.5\sigma$ -- $0.9\sigma$ (gold), than the ones of the $\Lambda$CDM, $\left<Z_\max\right> \approx 2.7\sigma$ -- $4.3\sigma$ (gray) and $w_0 w_a$CDM models, $\left<Z_\max\right> \approx 2.3\sigma$ -- $3.4\sigma$ (blue) for all three statistics. The means of the histograms, $\left<Z_\max\right>$, are higher than the $Z$-scores of the ground truths used in the simulations (vertical dashed lines) for all three models and three statistics. This is due to the look-elsewhere effect: when searching for a maximum deviation over a range of data points, noise becomes more likely to add up to an apparent signal, the more data points are included. This is most apparent for the coasting cosmology, where noisy realizations of $q(z) = 0$ with $Z_\true = 0$ can add up to a mean signal of $\left<Z_\max\right> \approx 0.5\sigma$ -- $0.9\sigma$. A similar, albeit mostly smaller, effect is observed for $\Lambda$CDM, $\left<Z_\max\right> -  Z_\true \approx 0.1\sigma$ -- $0.8\sigma$, and for $w_0w_a$CDM, $\left<Z_\max\right> - Z_\true \approx 0.1\sigma$ -- $0.5\sigma$. The look-elsewhere effect also leads to a smaller scatter of the simulations, $\std(Z) \approx 0.75$ -- $0.97 < 1$, compared to the uncertainties propagated from DESI data for all statistics. 
\begin{table}
  \centering
  \footnotesize
  \begin{tabular}{llcccccc}
    \toprule
    & & \multicolumn{3}{c}{$Z_{\rm max}$} & \multicolumn{3}{c}{$Z_{\rm avg}$} \\
    \cmidrule(lr){3-5}\cmidrule(lr){6-8}
    Stat. & Truth & $Z_{\rm true}$ & $\langle Z\rangle$ & $\std(Z)$ & $Z_{\rm true}$ & $\langle Z\rangle$ & $\std(Z)$ \\
    \midrule
    \multirow{3}{*}{$S_{\rm F}$}
      & Coasting     &0.00      &+0.85      &0.75      &0.00      &0.00      &1.00  \\
      & $\Lambda$CDM &+1.93      &+2.74      &0.75      &+3.27      &+3.27      &1.00  \\
      & $w_0w_a$CDM  &+1.84      &+2.33      &0.77      &+2.45      &+2.45      &1.00  \\
    \midrule
    \multirow{3}{*}{$S_V$}
      & Coasting     &0.00      &+0.49      &0.92      &0.00      &0.00      &1.00  \\
      & $\Lambda$CDM &+3.79      &+3.89      &0.96      &+3.53      &+3.53      &1.00  \\
      & $w_0w_a$CDM  &+2.79      &+2.86      &0.97      &+2.35      &+2.35      &1.00  \\
    \midrule
    \multirow{3}{*}{$S_{VH}$}
      & Coasting     &0.00      &+0.73      &0.82      &0.00      &0.00      &1.00  \\
      & $\Lambda$CDM &+3.97      &+4.30      &0.86      &+4.74      &+4.74      &1.00  \\
      & $w_0w_a$CDM  &+3.26      &+3.43      &0.90      &+3.17      &+3.17      &1.00  \\
    \bottomrule
  \end{tabular}
  \caption{Tests on MC simulations of the three statistics $S_F$, $S_V$ and $S_{VH}$. $Z_{\rm true}$ is the noiseless-truth value, $\langle Z\rangle$ is the mock-ensemble mean, and $\std(Z)$ is the standard deviation, for $Z_\max$ (Eq.~\ref{eq:Z_max}) and $Z_\avg$ (Eq.~\ref{eq:Z_avg}), respectively. If the universe were a realization of $\Lambda$CDM ($w_0w_a$CDM), one would expect to detect acceleration with $2.7\sigma$ -- $4.7\sigma$ ($2.3\sigma$ -- $3.4\sigma$), with $S_{VH}$ being the most sensitive test.}
  \label{tab:mock}
\end{table}

\begin{figure}[t]
    \centering
    \includegraphics[width=\columnwidth]{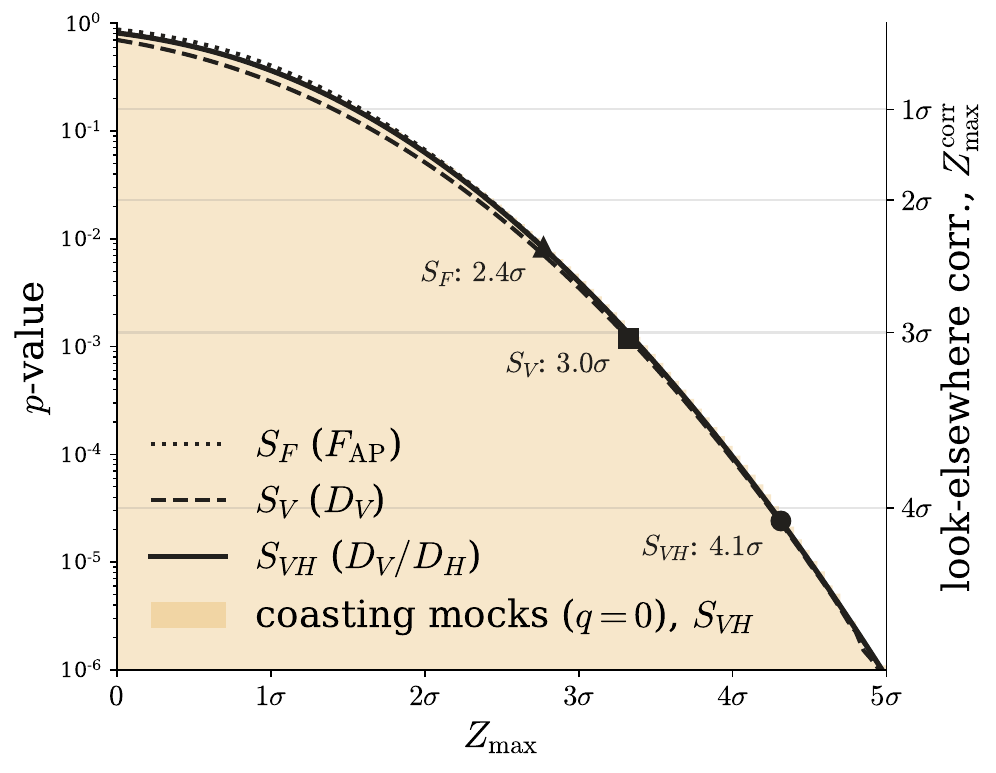}
    \caption{The survival probability ($p$-value) under a coasting null hypothesis, $p = P(Z_\max > Z | \mathrm{coasting})$, is obtained analytically (black lines). For comparison we show the histogram of the coasting mocks for $S_\mathrm{VH}$ (gold). We find $2.4\sigma$ -- $4.1\sigma$ evidence for acceleration among the three statistics (markers), when taking the look-elsewhere effect into account, $Z_\max^\corr$ (Eq.~\ref{eq:Z_max_corr}). 
    }
    \label{fig:Fig4}
\end{figure}

We account for the look-elsewhere effect by computing the $p$-value of $Z_\max$ under the null hypothesis of a never-accelerating universe. Since any never-accelerating universe has $S_\true(z_i) \le 0\ \forall z$, the coasting universe, $S_\true = 0$, is the limiting case closest to the data and its $p$-value bounds that of any never-accelerating model. Since all statistics are linear in the DESI measurements, the $Z_i$ are Gaussian with covariance $R_{ij} = [\Cov_S]_{ij}/[\sigma_S(z_i)\sigma_S(z_j)]$, and we can account for the look-elsewhere effect analytically. The maximum value of $Z_i$ (Eq.~\ref{eq:Z_max}) stays below the observed value $Z_\obs$ only if all $Z_i < Z_\obs$, so $p_\obs = 1 - \Phi_m(Z_\obs,\dots, Z_\obs;\, R)$, where $\Phi_m$ is the cumulative distribution function of an $m$-dimensional Gaussian with covariance $R_{ij}$. For $S_F$ the bins are independent and $p_\obs = 1 - \Phi(Z_\obs)^m$; for $S_V$ and $S_{VH}$ the shared BGS anchor induces correlations. We show the $p$-value for any given observed $Z_\max$ in Fig.~\ref{fig:Fig4}. The look-elsewhere-corrected significance is then
\begin{equation}
    \label{eq:Z_max_corr}
    Z_\max^\corr = \Phi^{-1}(1-p_\obs),
\end{equation}
where $\Phi$ is the cumulative distribution function of a 1-dimensional Gaussian. We checked that the analytic treatment agrees with the $p$-value obtained from the coasting MC simulations, where the latter is noisier due to the finite number of simulations in the tail.

The $Z$-score of real DESI data is indicated as the black solid lines in Fig.~\ref{fig:results} and quoted in the first column of Tab.~\ref{tab:results}. We find $Z_\max \approx 2.8\sigma$ -- $4.3\sigma$ with the highest evidence for the $S_{VH}$-statistic. These values are in good agreement with the histograms under $\Lambda$CDM and $w_0w_a$CDM and they represent a very unlikely realization of a coasting universe. Using the coasting universe as the null-hypothesis, we can derive a look-elsewhere-corrected $Z$-score via Eq.~\ref{eq:Z_max_corr}, which is shown as the markers in Fig.~\ref{fig:Fig4} and quoted in the middle column of Tab.~\ref{tab:results}. Taking the look-elsewhere effect into account we find that the evidence reduces by $0.25\sigma$ -- $0.38\sigma$ to $Z_\max^\mathrm{corr.} \approx 2.4\sigma$ -- $4.1\sigma$, again with the highest significance for $S_{VH}$.
\begin{table}[h]
  \centering
  \begin{tabular}{lccc}
    \toprule
    Statistic & $Z_\max$ & $Z_\max^\corr$ & $Z_\avg$ \\
    \midrule
    $S_F$ ($\Fap$)   & $2.77$ & $2.39$ & $2.19$ \\ 
    $S_V$  ($D_V$)  & $3.33$ & $3.04$ & $2.85$ \\ 
    $S_{VH}$ ($D_V/D_H$) & $4.31$ & $4.06$ & $3.22$ \\ 
    \bottomrule
  \end{tabular}
  \caption{Significance of acceleration from DESI DR2 as measured by the $Z$-score in units of $\sigma$ for the maximum deviation, $Z_\max$ (Eq.~\ref{eq:Z_max}), $Z^\corr_\max$ corrected for look-elsewhere effect (Eq.~\ref{eq:Z_max_corr}), and the average deviation, $Z_\avg$ (Eq.~\ref{eq:Z_avg}), for the three statistics.}
  \label{tab:results}
\end{table}

\subsection{Redshift-averaged significance, $Z_\avg$}
\label{sec:Z_avg}

Since acceleration presumably occurred over an extended redshift range, instead of picking the single bin with the largest deviation, we further test for acceleration by averaging over the first $m$ redshift bins:
\begin{equation}
    \label{eq:Z_avg}
    Z_\avg \equiv \frac{\bar S_m}{\sigma_{\bar S}} \quad \mathrm{with} \quad
    \bar S_m \equiv \frac{1}{m}\sum_{i=1}^{m} S(z_i).
\end{equation}
As for $Z_\max$ and described in App.~\ref{app:redshift}, we choose $m=3$. Since $\bar S_m$ is a linear combination of $S(z)$ for several $z$, its covariance is given by
\begin{equation}
     \sigma^2_{\bar S} = \frac{1}{m^2}\sum_{i=1}^{m}\sum_{j=1}^{m} [\mathrm{Cov}_S]_{ij}.
\end{equation}
We show the distribution of $Z_\avg$ in the bottom panels of Fig.~\ref{fig:results} and in Tab.~\ref{tab:mock}. The coasting universe has $\left<Z_\avg\right> \approx 0$, which is lower than for $\Lambda$CDM, $\left<Z_\avg\right> \approx 3.3\sigma$ -- $4.7\sigma$, and $w_0w_a$CDM, $\left<Z_\avg\right> \approx 2.4\sigma$ -- $3.2\sigma$. For every cosmology and statistic, $\left<Z_\avg\right>$ agrees with $Z_\true$ and $\std(Z) = 1$. This is expected since $Z_\avg$ is a linear combination of Gaussian variables, normalized by its own standard deviation, so it is Gaussian with mean $Z_\true$ and unit variance. Since no maximization over redshift bins is involved, there is no look-elsewhere effect to account for.

The $Z$-score of real DESI data (black solid lines in Fig.~\ref{fig:results}) lies at $Z_\avg \approx 2.2\sigma$ -- $3.2\sigma$ with the highest evidence for the $S_{VH}$-statistic. This is in agreement with both the $\Lambda$CDM and $w_0w_a$CDM simulations and in agreement with $Z_\max^\corr$ albeit at slightly lower significance.


\subsection{Discussion}
\label{sec:discussion}

From the three statistics used in this work, $S_{VH}$ is the most sensitive statistic. This is for two reasons, as already described in \cite{Sutherland:2011uy}. The first reason can be understood by considering what redshifts the different BAO measurements are sensitive to. While $D_H(z)$ measures $H(z)$ at $z$, $D_M(z)$ contains an integral from $0$ to $z$ (Eq.~\ref{eq:DH_DM}) and thus is sensitive to approximately $H(z/2)$, whereas $D_V(z)$ (Eq.~\ref{eq:DV}) is sensitive to $H(2z/3)$.\footnote{This can be seen by expanding $D_M$ around the midpoint $z/2$ and approximating the geometric mean for $D_V$ \cite{Sutherland:2011uy}.} Thus $S_{VH}$ compares $D_V(\zref)$ at the lowest redshift (here $\zref = 0.295$ for BGS, sensitive to $2\zref/3$) and $D_H(z)$ at a redshift around the transition redshift (here $z = 0.934$ for LRG3+ELG1, sensitive to $z$), and thus gives the largest lever arm. The second reason is that due to the angle averaging, $D_V$ has a factor $\sim 2.5$ smaller relative uncertainty than $D_H$ and $\sim 3.5$ smaller relative uncertainty than $\Fap$ (excluding the Ly-$\alpha$ bin). Note that LRG3+ELG1 also has the smallest relative uncertainty on $D_H$. Thus, even though $z_\mathrm{LRG3+ELG1} = 0.934$ is beyond the transition redshift at $z_\mathrm{t} \approx 0.6$ -- $0.9$, $Z_\max$ obtained from the simulations peaks there for $S_{VH}$ (App.~\ref{app:redshift}). For comparison, $\Fap$ compares $H(z/2)$ and $H(z)$ at the same tracer, and $S_V$ compares $H(2z/3)$ and $H(2\zref /3)$, providing smaller lever arms. Alternative statistics based on $D_H(z)/D_H(\zref)$, $D_M(z)/D_M(\zref)$ or $D_H(z)/D_M(\zref)$ are less sensitive than the statistics above due to a smaller redshift lever arm and/or larger uncertainties of $D_M$, $D_H$ compared to $D_V$. $S_{VH}$ thus balances redshift lever arm and measurement uncertainty. Note that while $S_F$ is less sensitive than $S_V$ and $S_{VH}$, it has the advantage that it is the only statistic that does not use BGS-$D_V$ as a reference; thus any systematic in BGS would bias $S_V$ and $S_{VH}$, but would not affect $S_F$.


\section{Conclusions}
\label{sec:conclusions}

In this work, we used model-independent and reconstruction-free tests for cosmic acceleration, assuming only homogeneity, isotropy, and BAO being a standard ruler. As a robustness check, we consider three different, albeit not independent statistics. Applied to DESI DR2 BAO data, all statistics find evidence for acceleration in $0 \leq z \leq 0.934$ when assuming flatness. We find the largest significance for the $S_{VH}$ statistic with  $Z_\max^\corr = 4.1\sigma$ for the look-elsewhere-corrected maximum point-wise significance and $Z_\avg = 3.2\sigma$ for the average significance over the three lowest redshift bins (with the BGS sample as reference). For the statistics $S_F$ and $S_V$ we find consistent but slightly lower significances with $Z_\max^\corr \approx 2.4\sigma$ -- $3.0\sigma$ and $Z_\avg \approx 2.2\sigma$ -- $2.9\sigma$. 

Allowing for spatial curvature does not change these conclusions (App.~\ref{app:curvature}). The sensitivity of the statistics degrades in different directions of $\Omega_k$: For closed cosmologies, $-1 \leq\Omega_k <0$, the significances $Z_\max^\corr(S_{VH}) \geq 3.5\sigma$ and $Z_\max^\corr(S_F) \geq 0$ degrade, while $Z_\max^\corr(S_{V}) =3.0\sigma$ stays unchanged. For open cosmologies, $0 <\Omega_k \leq1$, the significance $Z_\max^\corr(S_{V}) \geq 1.0\sigma$ degrades, while $Z_\max^\corr(S_F)=2.4\sigma$ and $Z_\max^\corr(S_{VH}) = 4.1\sigma$ stay unchanged. Thus for any $|\Omega_k|\le1$, at least two statistics retain evidence for acceleration, with the most sensitive one, $S_{VH}$, remaining above $3.5\sigma$ and the second-most sensitive above $2.4\sigma$.

Our work complements existing model-dependent constraints on acceleration from CMB, BAO and SNe and model-independent constraints from SNe. To our knowledge, all previous model-independent constraints on acceleration from BAO rely on a reconstruction between BAO data points, e.g.\ GPR, which introduces a dependence on the reconstruction method, especially with as little as six or seven BAO data points. The model-independent statistical tests in this work are inferred directly from the BAO data and do not require any reconstruction. Future BAO data sets, especially at low redshift, will improve the constraining power of the statistical tests for acceleration in this work.


\vspace{2mm}
\textbf{Acknowledgments:} LH was supported by a William H. Miller fellowship. The authors acknowledge the use of Anthropic’s Claude AI for code development, plotting, and manuscript reviewing. The scientific ideas, analysis choices, interpretation of the results, and manuscript writing are the responsibility of the authors. We also acknowledge the use of the Python packages \texttt{matplotlib}, \texttt{numpy}, \texttt{pandas}, \texttt{scipy}.


\appendix
\section{Curved cosmologies}
\label{app:curvature}

In this appendix, we relax the assumption of flatness. Allowing for non-zero curvature, $\Omega_k = -kc^2/(a_0 H_0)^2$, $k\in\{-1, 0, +1\}$, the comoving angular diameter distance becomes
\begin{equation}
    \label{eq:DM_curved}
    \begin{gathered}
        D_M(z) = \frac{c}{H_0}\, S_k\!\left(\chi(z)\right)\ \text{with}\\
        \chi(z) = \int_0^z\frac{\e^{-Q(z')}}{1+z'}\d z'\ \ \text{and} \\
        S_k(x)=\begin{cases}
            \sinh\!\big(\sqrt{\Omega_k}\,x\big)/\sqrt{\Omega_k}, & \Omega_k>0\ \text{(open)},\\
            x, & \Omega_k=0\ \text{(flat)},\\
            \sin\!\big(\sqrt{-\Omega_k}\,x\big)/\sqrt{-\Omega_k}, & \Omega_k<0\ \text{(closed)},
        \end{cases}
    \end{gathered}
\end{equation}
while $D_H$ and $E(z)$ are unchanged. 

For an open universe, $\sinh(x) \geq x$ and thus $S_k(x)\geq x$, so the lower bounds on $D_M$ used for $S_F$ and $S_{VH}$ (Sec.~\ref{sec:methods}) hold in the same way as for the flat case. $S_V$ instead requires an upper bound on $D_M(z)/D_M(\zref)$: since $\chi(\zref) \leq \ln(1+\zref)$, $S_k$ is monotonically increasing and $\sinh(Rx)/\sinh(x)$ increases with $x$ for $R\geq1$, this gives $D_M(z)/D_M(\zref) \leq S_k[\ln(1+z)]/S_k[\ln(1+\zref)]$. 

For a closed universe, $S_k$ is monotonically growing and concave, i.e.\ $\sin'(x)>0$ and $\sin''(x)<0$ for $x<\pi/2$, which holds at all DESI redshifts for $\Omega_k \geq -1$. This implies $\sin(rx) \geq r \sin(x)$ and thus $S_k(rx) \geq r\, S_k(x)$ for $0\le r\le1$. With $r=\e^{-Q}$, this gives $\Fap(z) \geq (1+z)\, S_k[\ln(1+z)]$ and the corresponding bound on $D_M(\zref)$ for $S_{VH}$, while $S_k(Rx)/S_k(x)\leq R$ leaves $S_V$ unchanged. 

This allows us to derive statistics for $\Omega_k \neq 0$. All statistics remain linear in the DESI measurements, so their uncertainties and the look-elsewhere correction (Eq.~\ref{eq:Z_max_corr}) carry over from the flat case.

\textbf{$\boldsymbol\Fap$-based statistic:} The generalized $S_F$-statistic follows from Eq.~\ref{eq:S_F}:
\begin{equation}
    \label{eq:S_F_k}
    \begin{gathered}
        S_F^{(k)}(z)\ \equiv\ (1+z)\,S_{k_-}\big(\ln(1+z)\big)-\Fap(z)\\
        \text{with}\ \ k_-\equiv\min(\Omega_k,0).
    \end{gathered}
\end{equation}
We show $S_F^{(k)}(z)$ from DESI data in the first three columns of Tab.~\ref{tab:curvature}. $S_F^{(k)}$ loses sensitivity only for $\Omega_k<0$, while the $Z$-scores for $\Omega_k>0$ stay unchanged compared to the flat case. Note that the limiting non-accelerating case of $S_F^{(k)}(z) = 0$ is not always a coasting universe: while for $\Omega_k \leq 0$, $S_F^{(k)}(z) = 0$ is reached for $q(z) = 0$; for $\Omega_k>0$ it is reached for  $Q(z) = \mathrm{const.} \rightarrow \infty$, i.e.\ a cosmology with $q(z) = 0$ for $z$ within the redshift range probed by DESI BAO but $q(z) \rightarrow \infty$ for $z\rightarrow 0$. Physically, the latter corresponds to a coasting universe at $z>0$ with a burst of deceleration today, $z \approx 0$. 

\begin{table}[t]
  \centering
  \resizebox{\columnwidth}{!}{%
  \begin{tabular}{rccccccccc}
    \toprule
    & \multicolumn{3}{c}{$S_F$} & \multicolumn{3}{c}{$S_V$} & \multicolumn{3}{c}{$S_{VH}$} \\
    \cmidrule(lr){2-4}\cmidrule(lr){5-7}\cmidrule(lr){8-10}
    $\Omega_k$ & $Z_\max$ & $Z_\max^\corr$ & $Z_\avg$ & $Z_\max$ & $Z_\max^\corr$ & $Z_\avg$ & $Z_\max$ & $Z_\max^\corr$ & $Z_\avg$ \\
    \midrule
    $-1.0$ & $0$ & $0$ & $0$ & $3.33$ & $3.04$ & $2.85$ & $3.80$ & $3.52$ & $2.66$ \\
    $-0.3$ & $1.32$ & $0.66$ & $0.80$ & $3.33$ & $3.04$ & $2.85$ & $4.16$ & $3.90$ & $3.05$ \\
    $-0.1$ & $2.29$ & $1.84$ & $1.72$ & $3.33$ & $3.04$ & $2.85$ & $4.26$ & $4.01$ & $3.16$ \\
    \hline
    $0.0$ & $2.77$ & $2.39$ & $2.19$ & $3.33$ & $3.04$ & $2.85$ & $4.31$ & $4.06$ & $3.22$ \\
    \hline
    $+0.1$ & $2.77$ & $2.39$ & $2.19$ & $2.93$ & $2.62$ & $2.57$ & $4.31$ & $4.06$ & $3.22$ \\
    $+0.3$ & $2.77$ & $2.39$ & $2.19$ & $2.14$ & $1.79$ & $2.02$ & $4.31$ & $4.06$ & $3.22$ \\
    $+1.0$ & $2.77$ & $2.39$ & $2.19$ & $1.43$ & $1.03$ & $0.11$ & $4.31$ & $4.06$ & $3.22$ \\
    \bottomrule
  \end{tabular}}
  \caption{Significance of acceleration from DESI DR2 allowing for curvature, $\Omega_k$, as measured by the $Z$-score in units of $\sigma$ for the maximum deviation, $Z_\max$ (Eq.~\ref{eq:Z_max}), $Z^\corr_\max$ corrected for look-elsewhere effect (Eq.~\ref{eq:Z_max_corr}), and the average deviation, $Z_\avg$ (Eq.~\ref{eq:Z_avg}), for the three generalized statistics $S^{(k)}_F$ (Eq.~\ref{eq:S_F_k}), $S^{(k)}_V$ (Eq.~\ref{eq:S_V_k}), $S^{(k)}_{VH}$ (Eq.~\ref{eq:S_VH_k}). The row $\Omega_k=0$ is the same as in Tab.~\ref{tab:results}. Negative $Z$-scores, where the data favor deceleration, are set to 0.
  }
  \label{tab:curvature}
\end{table}

\textbf{$\boldsymbol {D_V}$-based statistic:} The generalized $S_V$-statistic follows from Eq.~\ref{eq:S_V}:
\begin{equation}
    \label{eq:S_V_k}
    \begin{gathered}
        S_V^{(k)}(z)\ \equiv\ g_{k_+}(\zref)\, \frac{D_V(z)}{r_\mathrm{d}}-g_{k_+}(z)\, \frac{D_V(\zref)}{r_\mathrm{d}}\\
        \text{with}\ \ g_k(z) \equiv \left[\frac{z\, S_k^2(\ln(1+z))}{1+z}\right]^{1/3}\\
        \text{and}\ \ k_+\equiv\max(\Omega_k,0).
    \end{gathered}
\end{equation}
We show $S_V^{(k)}(z)$ from DESI data in the center three columns of Tab.~\ref{tab:curvature}. $S_V^{(k)}(z)$ only loses sensitivity for $\Omega_k > 0$, while the $Z$-scores for $\Omega_k<0$ stay unchanged, i.e.\ in the opposite direction compared to $S_F(z)$. Again, we observe that while for $\Omega_k >0$, $S_V(z) = 0$ is reached for $q=0$, for $\Omega_k <0$, the limit $S_V(z) = 0$ is reached for $Q(z) = \mathrm{const.} \rightarrow \infty$.  

\textbf{$\boldsymbol {D_V/D_H}$-based statistic:} The generalized $S_{VH}$-statistic follows from Eq.~\ref{eq:S_VH}:
\begin{equation}
    \label{eq:S_VH_k}
     S^{(k)}_{VH}(z) = (1+z)\, g_{k_-}(\zref)\, \frac{D_H(z)}{\rd} - \frac{D_V(\zref)}{\rd}.
\end{equation}
We show $S_{VH}^{(k)}(z)$ from DESI data in the three columns on the right of Tab.~\ref{tab:curvature}. $S_{VH}^{(k)}$ loses sensitivity only for $\Omega_k<0$, while the $Z$-scores for $\Omega_k>0$ stay unchanged compared to the flat case, i.e.\ in the same direction as $S_F^{(k)}$ but in the opposite direction as $S_V^{(k)}$. Analogously to $S_F^{(k)}$, for $\Omega_k \leq 0$, $S_{VH}^{(k)}(z) = 0$ is reached for $q(z) = 0$; for $\Omega_k>0$, it is reached for $Q(z) = \mathrm{const.} \rightarrow \infty$.

\begin{figure*}
    \centering
    \includegraphics[width=1\textwidth]{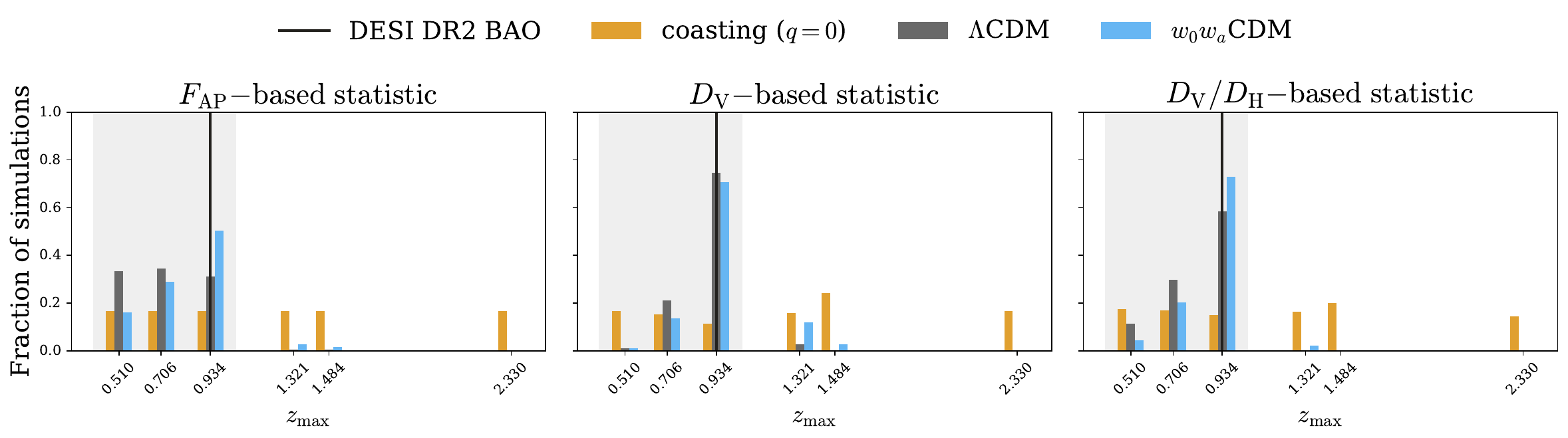}
    \caption{Distribution of the redshift, $z_\max$, that maximizes the acceleration signal, $Z_\max$ (Eq.~\ref{eq:Z_max}). Consistent with expectations, we find that the signal is largest in the first three redshift bins for $\Lambda$CDM and $w_0w_a$CDM (gray shaded area). We thus choose $m=3$ when quoting the maximum and average deviation from the non-accelerating universe in Sec.~\ref{sec:results}. $z_\max$ of DESI data (black vertical lines) agrees with the expectation from the accelerating cosmologies.}
    \label{fig:zmax}
\end{figure*}

When allowing for curvature, each statistic loses sensitivity in only one direction of $\Omega_k$: $S_F$ and $S_{VH}$ for closed ($\Omega_k<0$) and $S_V$ for open ($\Omega_k>0$) cosmologies. We consider $|\Omega_k|\leq1$, far beyond current cosmology-dependent constraints, e.g.\ within $\Lambda$CDM, $\Omega_k = 0.025 \pm 0.041$ from DESI alone and $\Omega_k = 0.0023 \pm 0.0011$ from DESI+CMB~\cite{DESI:2025zgx}. For $|\Omega_k|\leq1$, $S^{(k)}_{VH}$ is nearly unaffected, with $Z^\corr_\max \geq 3.5\sigma$. $S^{(k)}_F$ and $S^{(k)}_V$ are more sensitive to curvature, dropping to 0 at $\Omega_k = -1$ and $1.03\sigma$ at $\Omega_k=+1$, respectively. However, since they degrade in opposite directions of $\Omega_k$, at least one of them retains its flat-space significance, $Z^\corr_\max \geq 2.4\sigma$, for any $\Omega_k$. Thus, the evidence for acceleration persists when allowing for curvature, with the most sensitive statistic, $S_{VH}$, only slightly weakened compared to the flat case.


\section{Redshift distribution of $Z_\max$}
\label{app:redshift}

When reporting the significance for acceleration, we take into account the first $m$ redshift bins for $Z_\max$ (Eq.~\ref{eq:Z_max}) and for $Z_\avg$ (Eq.~\ref{eq:Z_avg}). We choose the maximum redshift bin $m$ informed by the MC simulations: we compute $Z_\max$ for each simulation, scanning over all redshift bins; the distribution of the redshift, $z_\max$, that maximizes $Z_\max$ is shown in Fig.~\ref{fig:zmax} for the three cosmological models described in Sec.~\ref{sec:data}.

For all three statistics, we find that $z_\max$ is within the first three redshift bins in $>97\%$ of simulations for $\Lambda$CDM and in $>85\%$ of simulations for $w_0w_a$CDM. This is expected for a transition from deceleration to acceleration at $z_t = 0.68$ for $\Lambda$CDM and at $z_t =0.85$ for $w_0w_a$CDM. Since all statistics contain integrals from 0 to $z$, evidence accumulates until $z_t$ and degrades beyond. The $\Lambda$CDM and $w_0w_a$CDM simulations peak at $z_\max=0.934$, which is slightly beyond $z_t$. This is due to the LRG3+ELG1 bin's smallest relative uncertainty and its long lever arm (c.f.\ Sec.~\ref{sec:discussion}). For the coasting cosmology, we find an approximately equal distribution among redshift bins consistent with the expectation of signal created by noise. Because the chance that noise alone produces a spurious signal grows with the number of bins included (look-elsewhere effect), we exclude the three highest-redshift bins. This becomes important when comparing DESI data to the null hypothesis of a coasting universe in order to correct for the look-elsewhere effect (Eq.~\ref{eq:Z_max_corr}). We thus choose $m=3$ for our analysis. For comparison, we show $z_\max$ of DESI data as black vertical lines, which lie at $z_\max = 0.934$ for all three statistics, in agreement with the $\Lambda$CDM and $w_0w_a$CDM simulations.


\bibliography{main}

\end{document}